\documentstyle[aps,multicol]{revtex}
\input{epsf}
\begin{document}
\draft
\title{Comment on ``The black hole final state''}
\author{Daniel Gottesman,$^1$  and John Preskill$^2$}
\address{$^1$ Perimeter Institute, Waterloo ON N2V 1Z3, Canada\\
$^2$ Institute for Quantum Information, California Institute of Technology,
Pasadena, CA 91125, USA}
%\date{{\bf PRELIMINARY DRAFT} -- 26 November 2003}
%% CALT-68-2466%%
\maketitle
\begin{abstract}

Horowitz and Maldacena have suggested that the unitarity of the black hole $S$-matrix can be reconciled with Hawking's semiclassical arguments if a final-state boundary condition is imposed at the spacelike singularity inside the black hole. We point out that, in this scenario, departures from unitarity can arise due to interactions between the collapsing body and the infalling Hawking radiation inside the event horizon. The amount of information lost when a black hole evaporates depends on the extent to which these interactions are entangling.
\end{abstract}
%\pacs{PACS numbers: ??}

%\narrowtext
\begin{multicols}{2}

%\section{Introduction}
Nearly 30 years ago, Stephen Hawking precipitated a crisis in quantum physics by discovering that black holes evaporate \cite{hawk1}. Hawking argued \cite{hawk2} that a process in which a pure quantum state collapses to form a black hole, which then evaporates completely, violates unitarity --- the final state of the emitted radiation is nearly thermal and therefore highly mixed. The crux of Hawking's argument is this: the geometry of the evaporating black hole contains spacelike surfaces that are crossed by both the collapsing body (inside the event horizon), and nearly all of the emitted Hawking radiation (outside the horizon). Therefore, {\em if} no quantum information is destroyed in the process, then the quantum state of the collapsing body must be ``cloned'' in the outgoing radiation. We infer, then, that either information is lost or cloning of arbitrary quantum states (which is inconsistent with the linearity of quantum mechanics) can occur; either way, we are pressed to accept that the foundations of quantum theory need revision.

One possible way to evade the conclusion that black holes destroy information is to adopt the principle of ``black hole complementarity'' \cite{complement}. One can decide not to be bothered by quantum cloning if it occurs only where no one can ever find out. Suppose that an observer stays outside the black hole long enough to verify that much of the collapsing body's quantum information is faithfully encoded in the Hawking radiation, and then dives into the black hole seeking confirmation that the collapsing body is still intact. Semiclassical reasoning does not suffice to answer whether she will succeed --- in order for the information carried by the collapsing body to reach the observer before meeting the singularity, it must be encoded in quanta with frequencies far exceeding the Planck frequency.

Thus liberated from a menacing semiclassical paradox, we may be satisfied provisionally to attain a consistent description of only the physics {\em outside} the horizon, in which information, rather than being lost, can be encoded in degrees of freedom localized on the horizon. There is strong evidence (especially from studies of asymptotically anti de Sitter spacetimes \cite{ads-cft}) that just such a description is provided by string theory. Indeed calculations providing a detailed quantitative picture of the microphysics of the event horizon arguably constitute string theory's greatest success.

The principle of black hole complementarity is useful, but it does not in itself fully resolve the puzzle of black hole information loss. For now we may be content with a consistent view of the world outside the horizon. But eventually, once a more complete understanding of quantum gravity becomes available, we should expect to be able to reconcile in detail the viewpoint of an observer who falls into a black hole with the viewpoint of an observer who stays outside.

In an interesting recent paper \cite{HM}, Horowitz and Maldacena
(HM) have proposed a quite simple way to reconcile the unitarity
of the black hole $S$-matrix with Hawking's semiclassical
reasoning. Their idea is to impose a boundary condition requiring
a particular quantum state at the black hole singularity. This proposal is attractive because it aims to move
the new physics associated with quantum gravity from the black hole
horizon, where one might expect to have an adequate semiclassical
description, to the singularity, where it is clear semiclassical
reasoning breaks down. Loosely
speaking, the intuition behind the HM proposal is this: to ensure
that no information is lost, we should leave no information behind
inside the black hole. The uniqueness of the final quantum state
at the singularity seems to prevent any information from getting
``stuck'' there, so that all of the information encoded in the
initial collapsing body can appear in the outgoing Hawking
radiation.  

The purpose of this comment is to point out that this intuition can be a bit misleading: it is not enough to impose a final state boundary condition; it is also important that the imposed final state be of a very special type.  For a particular natural decomposition of the quantum system inside the horizon into two parts, HM proposed that the final state is maximally entangled. Their scheme successfully restores unitarity if the two parts are noninteracting, or if the interactions are of just the right kind. But even weak interactions result in (weak) violations of unitarity.

Let us briefly review the HM proposal. The quantum state of the collapsing body belongs to a Hilbert space $H_M$ whose dimension is $N=e^S$, where $S$ is the black hole's entropy. In the semiclassical treatment of quantum field fluctuations on the background spacetime determined by the collapse and evaporation of the black hole, the Hilbert space of the fluctuations can be separated into two subsystems $H_{in}$ and $H_{out}$ (each also of dimension $N$) localized inside and outside of the horizon respectively. The (Unruh) quantum state $|\Phi\rangle_{in\otimes out}$ of the fields that looks like the vacuum in the far past is a maximally entangled pure state on $H_{in}\otimes H_{out}$
\begin{equation}
|\Phi\rangle_{in\otimes out}={1\over\sqrt{N}}\sum_{i=1}^N|i\rangle_{in}\otimes|i\rangle_{out}~,
\end{equation}
where $\{|i\rangle_{in}\}$ and $\{|i\rangle_{out}\}$ are orthonormal bases for $H_{in}$ and $H_{out}$ respectively. According to the HM proposal, the final state boundary condition imposed at the singularity requires the quantum state of $H_{M}\otimes H_{in}$ to be the maximally-entangled state
\begin{equation}
{}_{M\otimes in}\langle \Phi|(S\otimes I)~,
\end{equation}
where
\begin{equation}
{}_{M\otimes in}\langle \Phi|={1\over \sqrt{N}}\sum_{i=1}^N {}_M\langle i|\otimes {}_{in}\langle i |~,
\end{equation}
$S$ is a unitary transformation, and $\{|i\rangle_{M}\}$ is an orthonormal basis for $H_M$. The resulting transformation from $H_M$ to $H_{out}$ is
\begin{equation}
T\equiv {}_{M\otimes in}\langle \Phi|(S\otimes I)|\Phi\rangle_{in\otimes out}=\left({1\over N}\right)S~.
\end{equation}
The normalization factor $1/N$ indicates that if the quantum state of $H_{M}\otimes H_{in}$ were measured, the outcome prescribed by the HM boundary condition would occur with probability $1/N^2$. But under the terms of the HM proposal, all other measurement outcomes are to be discarded; the resulting state of $H_{out}$ is the ``postselected'' state under the assumption that the desired outcome is obtained. Renormalization of the postselected state removes the factor of $1/N$, and the transformation thus obtained is unitary --- no quantum information is destroyed in the process.

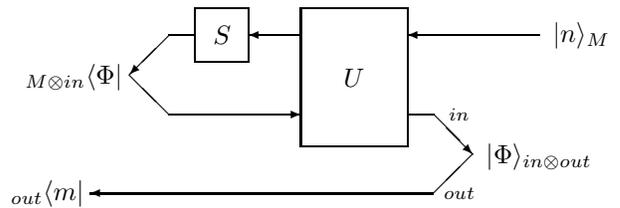
\begin{figure}
\begin{center}
\leavevmode
%\epsfxsize=6in
%\epsfbox{blackholecircuit.eps}
\begin{picture}(220,80)
%labels
\put(6,6){\makebox(0,0){${}_{out}\langle m|$}}
\put(16,50){\makebox(0,0){${}_{M\otimes in}\langle\Phi|$}}
\put(208,66){\makebox(0,0){$|n\rangle_M$}}
\put(192,20){\makebox(0,0){$|\Phi\rangle_{in\otimes out}$}}

\put(162,36){\makebox(0,0){${}_{in}$}}
\put(162,6){\makebox(0,0){${}_{out}$}}

\put(102,24){\framebox(40,52){$~U$ }}
\put(62,56){\framebox(20,20){$~S$ }}

\put(192,66){\vector(-1,0){50}}
\put(102,66){\vector(-1,0){20}}
\put(62,66){\line(-1,0){10}}

\put(52,66){\vector(-1,-1){15}}
\put(37,51){\line(1,-1){15}}

\put(52,36){\vector(1,0){50}}
\put(142,36){\line(1,0){10}}

\put(152,36){\vector(1,-1){15}}
\put(167,21){\line(-1,-1){15}}

\put(152,6){\vector(-1,0){130}}

\end{picture}

\end{center}
\caption{Information flow in the evaporating black hole, according to the Horowitz-Maldacena (HM) proposal. Here time runs from right to left, $|n\rangle_{M}$ is the initial quantum state of the collapsing body, and $|m\rangle_{out}$ is the final quantum state of the outgoing Hawking radiation emitted during evaporation.  The ket $|\Phi\rangle_{in\otimes out}$ is the maximally-entangled Unruh state of the infalling and outgoing Hawking radiation, and the bra ${}_{M\otimes in}\langle \Phi|$ is the (maximally-entangled) boundary condition imposed at the spacelike singularity. $S$ is the unitary black-hole $S$-matrix, which HM absorb into the boundary condition. The unitary matrix $U$, not considered by HM, arises from interactions between the collapsing body and the infalling radiation inside the event horizon.}
\label{fig:blackholecircuit}
\end{figure}

The flow of information from the collapsing body to the outgoing radiation, indicated in Fig.~\ref{fig:blackholecircuit}, can be described in the following somewhat fanciful language: The information propagates from past infinity to the black hole singularity. Rather than being absorbed there, it is ``reflected,'' propagating backward in time from the singularity to the preparation of the Unruh state. There it is reflected again, and propagates to future infinity.

In Fig.~\ref{fig:blackholecircuit}, we have also included a unitary transformation $U$ acting on $H_{M}\otimes H_{in}$. This transformation (not explicitly considered by HM) arises due to interactions of the collapsing body with the quantum field fluctuations after horizon crossing but before arrival at the singularity. Such interactions are certainly to be expected, and they blur the distinction between the two subsystems $H_{M}$ and $H_{in}$.

%\end{multicols}

%\begin{multicols}{2}

In our fanciful language, the transformation $U$ can be
interpreted as an interaction between the information's past and
future self. Various authors (Deutsch \cite{deutsch}, for example)
have pointed out that such interactions can cause a breakdown of
unitarity. Indeed, Schumacher and Bennett \cite{schumacher} have
observed that time travel can be simulated by combining quantum entanglement with postselection, and they have studied the departures from unitarity that result when interactions are also included. However, the case depicted in
Fig.~\ref{fig:blackholecircuit} differs slightly from that
considered in \cite{schumacher} because the two copies of the
information acted upon by $U$ propagate through time in opposite
directions.

With the unitary transformation $U$ included, the HM prescription yields the transformation
\begin{equation}
\langle m | T |n\rangle = {1\over\sqrt{ N}}\langle \Psi|n,m\rangle~,
\end{equation}
where
\begin{equation}
\langle \Psi |= \langle \Phi|(S\otimes I)U~.
\end{equation}
If $U$ is an arbitrary unitary transformation, then $\langle \Psi|$ is an arbitrary normalized pure state on $H_{M}\otimes H_{in}$, and therefore $T$ can be any matrix satisfying
\begin{equation}
\sum_{m,n} |\langle m | T |n\rangle|^2= 1/N~.
\end{equation}
If and only if $\langle\Psi|$ is a maximally entangled state, we recover the conclusion of HM, that the evolution governed by $T$ (after renormalization) is unitary. For any choice of $\langle \Psi |$ that is not maximally entangled, at least some information will be lost when the black hole evaporation is complete. If $\langle \Psi |$ is a product state, then the final state of the outgoing Hawking radiation will be independent of the initial state of the collapsing body, and the information loss will be complete.

For example, consider the case $(S\otimes I)U=V$, where $V$ is the controlled-sum gate that acts on an orthonormal basis according to
\begin{equation}
V\left(|i,j\rangle\right) = |i,j+ i~({\rm mod}~N)\rangle~
\end{equation}
(the first subsystem is the ``control'' and the second subsystem is the ``target'' of the gate); then
\begin{equation}
\langle m | T |n\rangle = {1\over\sqrt{ N}}\langle \Phi|V|n,m\rangle=\left({1\over N }\right)\delta_{m,0}~.
\end{equation}
The state of the outgoing radiation is $|0\rangle_{out}$, irrespective of the state of the collapsing body. If the direction of the controlled-sum were reversed (the control and target interchanged), then we would have
\begin{equation}
\langle m | T |n\rangle = \left({1\over N }\right)\delta_{n,0}~.
\end{equation}
In this case, the action of the state upon itself via the controlled-sum gate results in a ``time travel paradox'' unless the state of the collapsing body is $|0\rangle_M$.

The controlled-sum gate results in maximal violation of unitarity
(complete loss of information) because it is a maximally
entangling gate --- it can transform a product state to a
maximally-entangled state and vice versa. In general, the amount
of information loss is related to the entangling power of the
transformation $V=(S\otimes I)U$. Quantifying the amount of lost
information is an interesting open mathematical problem. The
matrix $T$ can be regarded as a kind of noisy quantum channel, but
it is not a trace-preserving completely positive linear map (the
type of channel whose capacity has been much studied by the
quantum information theorists --- see, for example \cite{shor});
rather the output is a {\em nonlinear} function of the input
because of the renormalization after postselection.

The interior of a black hole is a tumultuous place\cite{chaos}, where it is not easy to maintain a clear distinction between the two subsystems invoked in the HM proposal. In particular, far from the singularity, where the semiclassical description of the spacetime is still reasonably accurate, ordinary ``low-energy'' standard model interactions will entangle the collapsing body (consisting of pressureless dust, for example) with the infalling Hawking radiation (photons, for example). The degree of entanglement established in the semiclassical region may be small compared to the black hole entropy $S$, so one can imagine that, by slightly tweaking the boundary condition near the singularity to compensate for the interactions in the semiclassical region, unitarity of the black hole $S$-matrix can be salvaged. But this scheme seems to require an implausible conspiracy between low energy physics and Planck scale physics, which reduces the appeal of the HM proposal.

In the absence of such a conspiracy, one might reasonably propose
(because of the relatively weak entangling power of the
interactions in the semiclassical regime) that a final state
boundary condition leads to a small amount of information loss ---
that {\em much} information is hidden in subtle correlations among
the quanta emitted in the Hawking radiation, but not {\em all} of
the quantum information that was initially encoded in the
collapsing body. For us, this uncomfortable compromise has limited
appeal. Information loss, once allowed, tends to be highly
infectious, and it is difficult to formulate deformations of
quantum mechanics that incorporate a small amount of information
loss without detectable impact on low-energy experiments
\cite{banks}.

To summarize, the success of string theory in providing a microscopic understanding of black hole entropy still leaves unresolved the challenge of reconciling semiclassical reasoning with the putative unitarity of black hole evaporation. The HM proposal provides a potentially fruitful perspective on this problem, but leaves many mysteries unexplained.

\acknowledgments
We are grateful to Ben Schumacher and Charles Bennett for telling us about their work on ``conditional quantum time travel.''  We also thank Dave Bacon, Patrick Hayden, Gary Horowitz, and Juan Maldacena for helpful discussions and correspondence. This work has been supported in part by the Department of Energy under Grant No. DE-FG03-92-ER40701, by the National Science Foundation under Grant No. EIA-0086038, and by the Caltech MURI Center for Quantum Networks under ARO Grant No. DAAD19-00-1-0374.

\end{multicols}
\end{document}